\documentstyle[11pt,aaspp4]{article}

\slugcomment{Accepted for publication in ApJ Letters}

\newcommand\etal{{ et al. }}
\def\lsim{\mathrel{\rlap{\lower 4pt \hbox{\hskip 1pt $\sim$}}\raise 1pt \hbox
        {$<$}}}
\def\gsim{\mathrel{\rlap{\lower 4pt \hbox{\hskip 1pt $\sim$}}\raise 1pt \hbox
        {$>$}}}

\begin{document}

\righthead{Relativistic Bullets 
Ejection from Supernovae and Generation of Gamma
Ray Bursts}

\title{Relativistic Bullets 
Ejection from Supernovae and Generation of Gamma
Ray Bursts}

\author
{Hideyuki UMEDA}
\affil{ Research Center for the Early Universe, and
Department of Astronomy, 
School of Science, University of Tokyo, \\ Bunkyo-ku, 
Tokyo 113-0033, Japan \\E-mail: umeda@astron.s.u-tokyo.ac.jp}

\begin{abstract}
 It is generally believed that cosmological Gamma Ray Bursts
(GRBs) are produced by the deceleration
of relativistic objects with $\Gamma \gsim 100$.
 We study the possibility that some GRBs are produced
along with relativistic matter ejection from supernovae. 
In this model,
it is quite likely that the matter has to travel through
the progenitor's thick envelope before generating GRBs.
Under the assumption
that the ejected matter is described as a single collective matter,
we obtain constraints on the matter
to have $\Gamma \gsim 100$ at the breakout of the progenitor.
One advantage of considering this type of model is that the expected
GRB energy is sufficiently large, in contrast to the 
GRB generation model by the shock breakout in the energetic
supernova explosion.
We find that in general the cross section of the matter has to be 
very small compared with the progenitor's radius
and thus the matter has to be bullet (or jet)-like
rather than shell-like.
\end{abstract}


\section{Introduction}

 It is generally believed that cosmological Gamma Ray Bursts
(GRBs) are produced 
by relativistic matter with Lorentz factor $\Gamma \gsim 100$
when the kinetic energy of the matter is converted into radiation
by inelastic collision with other relativistic matter --- internal
shock model ---, or by collision with interstellar matter (ISM) --- 
external shock model (e.g. Piran 1999).

 Although it is still controversial, there are indications
that some GRBs might be associated with type Ib/c supernovae
(SNe) (e.g., Galama \etal 1998, Bloom \etal 1999, Castro-Tirado \&
Gorosabel 1999, Reichart 1999). 
Especially a type Ic SN, SN1998bw, which
was discovered in the error box of GRB980425 has been considered to be 
the most promising candidate for the GRB-SN 
association (Galama \etal 1998, Iwamoto \etal 1998, Nakamura 1999). 
In Iwamoto \etal (1998) we considered the following scenario for the
GRB formation: at the shock breakout in the energetic explosion,
the surface layer becomes highly relativistic, and the interaction 
with ISM may generate a GRB. One difficulty with this kind of 
model was that
the total energy stored in the relativistic region was very small, 
and hence inconsistent with the observed GRB energy (see also 
Woosley, Eastman \& Schmidt 1999). Therefore, something 
more than the energetic explosion is necessary to produce a GRB 
from a SN explosion. This led to the investigation of jet-like explosion
(Khokhlov \etal 1999) or the effects of disk accretion around a 
black hole (BH) (MacFadyen \& Woosley 1999).

 In this {\it Letter}, we assume that during or after core collapse,
a small fraction of the matter acquires a large energy, 
becomes relativistic 
deep inside the SN progenitor, and is ejected. This type of matter
ejection may be related to the "mystery spot" of SN1987A
(e.g.,  Rees 1987, Piran \& Nakamura 1987, Cen 1999).
We show that such relativistic
matter can generate sufficiently energetic GRBs if the cross section
of the matter is  sufficiently small.

 The collision of the relativistic matter 
with the surrounding matter produces an external shock, 
which decelerates the matter.
If the relativistic matter is in an optically thin region,
prompt radiation from it may be observed as a GRB.
This kind of situation may
be realized, if the collapsed star is rapidly rotating and forms 
an accreting BH, and if matter is sufficiently evacuated along 
the rotational axis  as in the model of Fryer \& Woosley (1998).
However, numerical simulations of such system 
 (MacFadyen \& Woosley 1999) suggest that
although the density is significantly reduced  near the central BH
along the polar direction, it is still high. 
Also the central BH and accretion disk are still surrounded by the
envelope of the progenitor. Furthermore,
light curve modeling of SN1998bw suggests that the SN explosion
was globally well described by the usual SN model, i.e., the 
explosive SN shock should propagate through the normal stellar 
envelope (Iwamoto \etal 1998, Woosley \etal 1999). Therefore, it is likely
that the relativistic object produced near the center
has to travel through the optically thick envelope of the progenitor.
The key question is, thus, if the object can still have
$\Gamma \sim 100$, when it breaks out of the envelope.
By considering a simple model, we obtain the constraints
for a relativistic object produced in the deep interior 
of a SN progenitor to
have $\Gamma \simeq 100$ when it breaks out of the envelope.

\section{Basic equations}

 Here, the basic equations and some solutions
are shown to describe the slowing down of a
relativistic object via collision with the surrounding matter.
 
 We assume that the slow-down of the relativistic object 
occurs by a series of infinitesimal 
inelastic collisions between the object and infinitesimal masses.
We denote the rest-frame energy (rest-mass and
thermal energy) of the object located at radius $r$
by $M(r)$, its Lorentz factor by $\Gamma(r)$, the mass of the envelope
that has already collided with the object by $m(r)$, and the thermal
energy produced in the collision by $E$. Then the energy and momentum
conservation yield (e.g., Piran 1999), 
\begin{equation}
\frac{d\Gamma}{\Gamma^2-1}=-\frac{dm}{M}, \quad
{\rm and}~ \quad dE=(\Gamma-1) c^2 dm.
\end{equation}
The variation of $M$ is given by $dM =(1-\epsilon) dE/c^2+dm$, where
$\epsilon$ is the fraction of the shock-generated thermal energy that
is radiated. 
When the object is in the envelope, we assume $\epsilon \sim 0$.
If $\Gamma_i \gg 1$ and $\Gamma \gg1$ these equations can be easily
integrated to give:
\begin{equation}
\label{mr1}
 M(r) \simeq M_0 \left(\frac{\Gamma_i}{\Gamma(r)}\right)^{1-2\epsilon},
\end{equation}
where $M_0$ and $\Gamma_i$ are initial rest-mass and Lorentz factor,
and 
\begin{equation}
\label{mr2}
 m(r) \simeq \frac{M_0 \Gamma_i}{2-\epsilon} 
(\Gamma(r)^{-2+\epsilon}-\Gamma_i^{-2+\epsilon}).
\end{equation}

 After breaking out from the envelope, the object emits radiation as
it is slowed down. If the collisions between relativistic
objects are not efficient, then slow-down occurs mostly via
collision with ISM (external shock model). 
Eq.(\ref{mr2}) shows that, by the time the object lose
half of its kinetic energy ($\Gamma \simeq \Gamma_i/2$), the mass of
$m_s \sim m(R)/\Gamma$ has been plowed. The total thermal energy
produced is  (Meszaros \& Rees 1992), 
\begin{equation}
\label{etot}
E \simeq \int_0^{m_s} \Gamma c^2 dm \sim m(R)c^2,
\end{equation}
which is assumed to be roughly equal
to the total GRB energy.

 In the external shock model, the characteristic synchrotron energy is
given by $h\nu =$ 160 keV $\epsilon_B^{1/2} \epsilon_e^2
 (\Gamma/100)^4 n^{1/2}$, where $\epsilon_B$ is the ratio of
the magnetic field energy density to the total thermal energy,
$\epsilon_e$ is the fraction of the total thermal energy which goes
into random motions of the electrons, and $n$ is the ISM 
number density per cm$^3$ (e.g. Piran 1999). 
Therefore, $\Gamma \gsim 100$ is required
for efficient gamma ray emission. 

\section{Conditions for GRB generation}

 We assume that the objects become relativistic
near the collapsed core, and travel through the envelope
of the SN progenitor.
 In the following, we consider two cases for the deceleration
of the objects.
One is when the relativistic objects do not
lose their energy significantly in the envelope of the
progenitor, i.e.,
$\Gamma_i \simeq \Gamma_f $, where $\Gamma_f$ is the
Lorentz factor at breakout. 
The second case is when the 
objects are initially highly relativistic and slow down to 
$\Gamma_f \ll \Gamma_i$ through the envelope. 
In this paper, for simplicity, we only consider the case with
$\Gamma_f \simeq  100$. 
 In both cases we find that the cross sections of the relativistic
objects must be very small compared with progenitor's radius.
Hence the object should be bullet
(or beam or jet)-like, rather than shell-like.
Note that Eq.(1) applies only when the object can be treated 
as collective matter, and hence it does not apply
to long beams or jets.

\subsection{Case I: $\Gamma_i \simeq \Gamma_f \simeq 100$}

 The energy $E$ required to accelerate $N'$ pieces of bullets with 
mass $M_0$ up to $\Gamma_i$ is $E=N' M_0 c^2 \Gamma_i$. Therefore,
\begin{equation}
\label{m0g}
 N' M_0 = 5.5\times 10^{-4} E_{51} \Gamma_i^{-1} (M_\odot),
\end{equation}
where $E_{51} \equiv E / (10^{51}$ erg). As shown in the previous section,
when each bullet plows a mass of 
$\sim M_0/\Gamma$, $\Gamma$ becomes roughly half. 
Therefore, if the bullet 
is not to be significantly decelerated after traversing
a mass $M_{env}$ and radius $R$,
the diameter of the bullets $d_b$ (here the diameter is assumed
to be constant) should satisfy the condition:
 $d_b < ( 4\pi M_0/3 M_{env} \Gamma_i)
 ^{1/2} R$.
For example, for the progenitor model of SN1998bw, $M_{env} 
\sim 10M_\odot$
(Iwamoto \etal 1998), using Eq.(\ref{m0g}), we obtain
\begin{equation}
\label{dbc1}
d_b < 10^{-4} \sqrt{E_{51}/N'} (100/\Gamma) R.
\end{equation} 

 The situation considered here is similar to the model considered 
in Heinz \& Begelman (1999). They assume that the central 
engine of the burst
releases $N'$ bullets distributed over an opening angle of $\theta
\sim 10^{\circ}$ with $\Gamma_i \sim 1000$. Each bullet is assumed to be
expanding sideways with
a velocity $v_s = \alpha c /\Gamma \ll c \Gamma$ measured in the
observer's frame ($\alpha c$ is the expansion velocity in the 
comoving frame).
Since the viewing angle $1/\Gamma \ll \theta$, the
observed number of bullets
is $N \simeq N'/(\theta^2 \Gamma^2)$. 
They considered this model to explain the
short-time variability of the canonical GRBs, and claimed that
the external shock model can explain that variability
if $N\sim 100, \alpha \sim 0.01, \Gamma \sim 10^3$, and the ambient gas
density $n \sim 10^8$cm$^{-3}$. In this paper we do not attempt
to produce short time variability, since it is not certain 
that a GRB associated with a SN should show it. 

 Each bullet may expand if it has sufficiently large internal
energy, though the expansion rate depends on several assumptions.
If we assume that the bullets expand sideways with comoving
velocity $\alpha c$, as in Heinz \& Begelman (1999), then condition
(\ref{dbc1}) is somewhat relaxed as shown in the following. 
The cross section of the bullet in the observer's frame
$S(r)$ is given by
  $S(r)\simeq  (2 v_s t \Gamma^2)^2 = (\alpha r/\Gamma)^2$,
where $v_s =\alpha c/\Gamma$ is the sideways velocity in the
observed frame, and
$t \simeq r/(2c \Gamma^2)$ is the observer's time when
the bullet is located at radius $r$.
The total mass plowed by the bullet $m_s$ is given by,
\begin{equation}
\label{ms1}
 m_s = \int_{r_0}^{R} \rho(r) S(r) dr,
\end{equation} 
where $R$ is the radius of the progenitor, $r_0 (\ll R)$  is the
radius where the bullet is accelerated to $\Gamma=\Gamma_i$, 
and $\rho(r)$ is the density of the progenitor.

 If $\rho \propto r^{-b}   (b < 3)$, and $r_0 \ll R$,
Eq.(\ref{ms1}) yields 
 $m_s \simeq (4\pi)^{-1} M_{env} \alpha^2 \Gamma^{-2}$, where
$ M_{env}$  
$\equiv  \int_{r0}^{R} 4\pi r^2 \rho(r) dr$ is the progenitor
mass above radius $r_0$. Note that $m_s$ is independent of $b$.
A constraint on $\alpha$ is given by $m_s < M_0/\Gamma$:
\begin{equation}
\alpha < (4\pi \Gamma M_0/M_{env})^{1/2} =
 0.08 \sqrt{\frac{E_{51}}{N' M_{env}(M_\odot)}}.
\end{equation}
The diameter of the bullet at breakout $(r=R)$, $d(R)$ is
\begin{equation}
\label{drc1}
d(R) = \alpha R/\Gamma < 10^{-3} \sqrt{\frac{E_{51}}
{N' M_{env}(M_\odot)}} R.
\end{equation}

 After the bullet breaks out of the envelope, a GRB is produced.
If it is produced
by the external shock, total energy of the order of $E_{GRB} 
\sim M c^2$ is released as the bullet plows through a mass of 
$\sim M/\Gamma_f$. Here $M$ is the rest-energy of the bullet at 
$r=R$, which is estimated by
Eq.(\ref{mr1}), $M \simeq M_0 \Gamma_i/\Gamma_f \simeq M_0$.
Using Eq.(\ref{etot}), we obtain the total GRB energy
$E_{GRB} \sim 10^{49} E_{51} (100/\Gamma_f) (N/N')$ erg.
 Note that if many bullets ($N'\gg 1$) are ejected within the opening
angle $\theta$, the total GRB energy seen by an observer is reduced
by the factor of $N/N' \simeq 1/(\theta \Gamma)^2$.

\subsection{Case II: $\Gamma_i \gg \Gamma_f \simeq 100$}

 Suppose again that the bullets expand sideways with constant velocity
$\alpha$ measured in the comoving frame. Then the diameter of the
bullet at radius $r$ in the observed frame is given by,
\begin{equation}
\label{dr2}
 d(r)= \int_{r_0}^{r} \alpha/\Gamma(r') dr'.
\end{equation}
In general $\Gamma(r)$ is a function of $\alpha$ and $\rho(r)$.
However, here we simply assume that 
$\Gamma(r)=\Gamma_i (r/r_0)^{-a}$ and $\rho(r)=\rho_0 (r/r_0)^{-b}$  
in order to get a rough constraint without obtaining an
exact solution for $\Gamma(r,\alpha, \rho(r))$. 
Here the subscript 0 represents 
the quantity at $r=r_0$ in the observer's frame. 
Strictly speaking, assuming that both $a$ and $b$ are constants is not
self-consistent, however the result given in Eq.(\ref{drc2}) is
not so sensitive to the value of $a$ and $b$. Therefore, 
this assumption is
sufficient for a qualitative analysis.
Now Eq. (\ref{dr2}) gives
$d \propto r^{1+\alpha}$. Hence we denote 
$ d(r)= d_0 (r/r_0)^{1+a}$,
where $d_0 = \alpha r_0/[(a+1)\Gamma_i]$.
If constants $a$ and $b$ satisfy $0<b <3$, and $2a+3-b>0$, then 
Eqs.(\ref{mr2}) and (\ref{ms1}) give
\begin{equation}
\label{m0}
 m_s \simeq  \frac{d_0^2 r_0 \rho_0}{2a+3-b}
 \left(\frac{R}{r_0}\right)^{2a+3-b} \simeq \frac{\Gamma_i}{2\Gamma_f^2}.
\end{equation}

We define the envelope mass by $M_{env} \equiv \int_{r_0}^{R} 
4\pi r^2 \rho(r') dr' \simeq $
$ (4\pi r_0^3 \rho_0 /3-b) (R/r_0)^{3-b}$.
Now, the diameter of the bullet $d$ in the observed frame 
at the progenitor surface ($r=R$) is given by 
\begin{equation}
\label{drc2}
 d(R)=d_0 \left(\frac{R}{r_0}\right)^{1+a} = 
\frac{R}{\Gamma_f} \sqrt{\frac{2a+3-b}{3-b} 
\frac{2\pi M_0 \Gamma_i}{M_{env}}} \simeq  \frac{0.06 R}{\Gamma_f} 
\sqrt{\frac{2a+3-b}{3-b} \frac{E_{51}}{N' M_{env} (M_\odot)}},
\end{equation}
or
\begin{equation}
 \alpha= 0.06 (a+1)
\sqrt{\frac{2a+3-b}{3-b} \frac{E_{51}}{N' M_{env} (M_\odot)}}.
\end{equation}
Note that 
if $\alpha$ is significantly smaller 
than the right hand side of this equation, Case I applies.
This result is not 
sensitive to the values of $a$ and $b$ 
as long as $2a+3-b/3-b \sim O(1)$. This justifies our
approximate analysis assuming that $a$ and $b$ are constants. The
total GRB energy (in the external shock model) can be estimated as 
$E_{GRB} \sim 10^{49} E_{51} (100/\Gamma_f) (N/N')$ erg, which
is same as Case I.

\section{Discussion }

 We have investigated relativistic matter ejection from 
SN progenitors in order to study the possibility that SNe
generate GRBs. If we consider this type of
model, it is likely that the relativistic matter has
to travel through the optically thick envelope of the SN progenitor
before generating the GRB.
 We have obtained conditions for such relativistic matter to 
have $\Gamma \gsim 100$ after passing through the thick
envelope of the progenitor. In general, in order to satisfy these
conditions, the cross section of the objects must be small. Typically
the diameter of the objects must be at least factor of
$10^3 E_{51}^{-1/2}$ smaller than the radius of the envelope when
they break out of the envelope. Therefore, the objects should 
be bullet (or beam or jet)-like rather than shell-like. 
Here we should note that the opening angle of GRB can be significantly
larger than the size of each bullet at breakout, because
the size of the bullet
may increase significantly by the time of GRB emission
owing to its expansion or to merging with other bullets.

 We make no attempt to explain how such bullets may be produced 
during core collapse since the explosion mechanism of massive
stars are not well understood. Indeed, currently
no mechanism is known to make such bullets.
 It is only speculation but such
bullets may be produced if the exploding core is 
inhomogeneous (e.g. Burrows \& Hayes 1996) 
and the explosion energy is concentrated in some small
regions, or if the collapsing core forms an accreting black hole
and relativistic bullets or jets are produced around the 
black hole (a jet of short duration may look like a bullet).
 
 Next let us estimate the initial rest-frame density of the bullets.
 Suppose the bullet has a volume of roughly $d_0^3$ in the 
observed frame, and has 
$\Gamma= \Gamma_i$ at mass coordinate $M_r =2M_\odot$ in
a SN progenitor.
If we take a progenitor model of SN1998bw, at $M_r =2M_\odot$
the radial coordinate is $r_0 = 2\times 10^8$cm. 
From Eqs. (\ref{drc1}) and
(\ref{drc2}), we find that
$d_0 \simeq 10^{-1} \Gamma_a^{-1} 
(E_{51}/N' M_{env} (M_\odot))^{1/2} r_0$ if we require
 $\Gamma_f \gsim 100$. Here $\Gamma_a >100$ for
Case I and $\Gamma_a=\Gamma_i>100$ for Case II. The bullet mass
is $M_0 \sim \Gamma_i d_0^3 \rho_i$. Thus, using 
Eq.(\ref{m0g}), one obtains a constraint
$ \rho_i \gsim 10^8 \Gamma_a (N'/E_{51})^{1/2}
 M_{env}^{3/2}(M_\odot) ({\rm g cm}^{-3})$.
This density is much larger than the progenitor's density
at this radius $\rho(r_0) \sim 10^8 $g cm$^{-3}$, which means
that the bullets must be produced deeper inside. 

The GRB energy produced by each bullet is 
$E_{GRB} \sim 10^{49} E_{51} N'^{-1} (100/\Gamma_f)$ erg (in the external
shock model). If $E_{51} \lsim 1$, this is sufficiently
large and not too large
to explain the energy of GRB980425 ($\sim 10^{48} f_b^2$ 
erg, if its distance is $\sim 40$Mpc,
where $f_b\leq 1$ is the beaming factor) 
which might be associated with SN1998bw. 

 The GRB980425 lasted about 30 seconds and this must
be explained in our model, if it is produced by bullets 
ejected from the progenitor.
The time scale of the GRB in this model is determined by the time
scale of the deceleration of the bullets, or
by the life time of the central engine. If it is determined by 
the latter, we can assume anything because almost nothing is 
known about the central engine. In the following, assuming the
external shock model, we discuss how
the time scale is determined in the former case.

 In our model,  the bullet mass at breakout is
$M \simeq M_0 \Gamma_i/\Gamma_f$. Each bullet emits a
GRB as it plows ISM mass of $m_s \sim M/\Gamma_f$. The mass plowed
 depends on ISM density and on the bullet's cross section.
If the ISM is produced by wind mass loss from the progenitor,
the number density can be estimated by $n_{ISM} (r)= 1.5\times 10^{36}
\dot M_{-6} v_{20}^{-1} r^{-2}$ cm$^{-3}$, where $\dot M_{-6}$
is the mass loss rate in units of $10^{-6}M_\odot yr^{-1}$ and
$v_{20}$ is the wind velocity in units of 20km s$^{-1}$. 
$\dot M_{-6}/v_{20}$
is roughly $\sim 0.1$ for O stars and $\sim 1 - 100$ for 
red giant stars.  The time evolution
of the bullet size is uncertain. The radius may increase
due to internal expansion or by merging with other bullets.  
Here we assume that the diameter of a bullet increases as 
$\propto r^{1+k}$ with $k \sim O(1)$. 
Suppose the diameter of the bullet is
$\simeq 10^{-2} (E_{51}/N')^{1/2} \Gamma_f^{-1} R$ 
at $r=R$, where $R$ is the
radius of the order of the progenitor radius. 
For the model of SN1998bw progenitor, $R \sim 10^{11}$cm.
Then the time scale of the GRB duration, $\tau$, is estimated by
\begin{equation}
\tau = \frac{\Delta R}{2\Gamma_f^2 c} \sim 
 \frac{R_{11}}{\Gamma_f^2} \left(\frac{10^{11} }
{\dot M_{-6} v_{20}^{-1} R_{11}} \right)^{\frac{1}{1+2k}}
{\rm sec},
\end{equation}
where Eqs. (\ref{m0g}) and (\ref{ms1}) are used, 
$\Delta R$ is the distance the bullet travels 
during GRB emission, and $R_{11}\equiv R/10^{11}$cm.
This result shows that if $k=0$, $\tau$ is independent of
$R$ and typically too long: $\tau > 10^5 (100/\Gamma_f)^2$ sec.
On the other hand, if $k \sim 0.5$,
$\tau \sim 30$ sec is possible.
If $k$ is larger, even a milli-second time scale may be possible.
Therefore, we may conclude that GRB durations of milli-seconds
to months may be realized with this model, depending on the parameters
and situations. Especially, the expansion rate of the bullet radius is
very important to determine the time scale. In this $Letter$
the expansion rate was treated as a free parameter, and constrained
it for successful GRB generation. However, the rate may be 
estimated if the pressure inside and outside the bullets
is given. The evolution of bullets including their expansion rate 
will be further explored by performing hydrodynamical simulations
in future work.

 Finally we comment on the internal shock model for the GRB
generation. In the last paragraph, we mentioned that the
cross section of the bullets may be increased via merging with
other bullets. If this is the case, the GRB may be produced by the
internal shocks produced during merging. If the merging bullets
grow larger than the angular diameter of $\Gamma^{-1}$ before losing
their kinetic energy, their subsequent evolution
will be similar to the model by Kumar \& Piran (1999). 
They considered
collision of relativistic blobs with total kinetic energy 
10$^{52}$ erg as the source of GRBs, and showed that the observed 
diversity of the GRB energy can be explained by the inhomogeneity
of the blobs and the differences of lines of sight. 
In general, the efficiency
of the kinetic to thermal energy conversion is much worse in
the internal shock model than in the external shock model. However,
the time scale of energy loss can be several orders of magnitudes smaller
than the external shock model and this may be an advantage
(e.g. Piran 1999). 

This work has been 
supported in part by the grant-in-Aid for
Scientific Research (0980203) of the Ministry of Education, 
Science, Culture and Sports in Japan.

\end{document}